\documentstyle[amssymb,aps,multicol]{revtex}

\begin{document}
\preprint{}
\draft

\title{Dynamical Decoupling of Open Quantum Systems}
\author{Lorenza Viola${}^1$, Emanuel Knill${}^2$, and Seth 
Lloyd${}^{1}$ \thanks{ 
Electronic addresses: vlorenza@mit.edu; knill@lanl.gov; 
slloyd@mit.edu.} }
\address{ ${}^1$ d'Arbeloff Laboratory for Information Systems and 
Technology, 
Department of Mechanical Engineering, \\ Massachusetts Institute of 
Technology, 
Cambridge, Massachusetts 02139 \\
${}^2$ Los Alamos National Laboratory, Los Alamos, New Mexico 87545} 

\maketitle

\begin{abstract}
We propose a novel dynamical method for beating decoherence and 
dissipation in open quantum systems. We demonstrate the possibility of 
filtering out the effects of unwanted (not necessarily known) 
system-environment interactions and show that the noise-suppression procedure  
can be combined with the capability of retaining control over the effective
dynamical evolution of the open quantum system.
Implications for quantum information processing are discussed.

\end{abstract}


 
\begin{multicols}{2}

All real world quantum systems interact with their surrounding
environment to a greater or lesser extent. Such systems are said to be
{\sl open}.  No matter how weak the coupling that prevents the system from
being isolated, the evolution of an open quantum system is eventually
plagued by nonunitary features like decoherence and dissipation
\cite{gardiner}.  Quantum decoherence, in particular, is a
purely quantum-mechanical effect whereby the system loses its ability
to exhibit coherent behavior by getting entangled with the ambient
degrees of freedom. Decoherence stands as a serious 
obstacle common to all applications relying on the capability
of maintaining and exploiting quantum coherence. These encompass
quantum state engineering \cite{eng}, quantum interferometry \cite{interf},
macroscopic quantum mechanics \cite{mqm} and, notably, the whole emerging 
field of quantum information processing \cite{qcomp}.

Recently, considerable effort has been devoted to design strategies able to
counteract the effects of environmental couplings in open-system evolutions.
In particular, the theory of quantum error correction has been developed to 
meet this challenge \cite{qecc}. From a physical point of view, the general
underlying question can be stated in terms of attaining {\sl quantum 
noise control}. Unlike the closed-system limit, where the active
manipulation of unitary dynamics is currently realized to be a problem
of quantum control theory \cite{control}, the possibility of {\sl
directly} exploiting control techniques to influence open-system properties
has not been fully explored yet. Existing approaches mainly rely on 
feedback (or closed-loop) control configurations \cite{milburn}. 
In fact, conventional quantum error correction protocols can be 
regarded, in their essence, as a form of quantum feedback control 
implemented on a {\sl redundant} physical system.

In this Letter, we formulate a model for {\sl decoupling a generic 
open quantum system from any environmental interaction} through simpler 
so-called open-loop control techniques. We show that the resulting 
description not only provides a comprehensive framework for 
decoherence-suppression schemes as first identified in \cite{viola} and
subsequently implemented by various authors under specific assumptions 
\cite{duan}, but, in contrast to previous proposals, it also points out 
a general criterion for engineering
{\sl effective} open-system evolutions that are, in principle, immune 
to noise and decoherence. More precisely, we find that one can effectively 
control the system to undergo a wide range of dynamical behavior while still 
eliminating the effects of the environment. The allowable dynamics are 
generated by a {\sl subgroup} of the possible system tranformations.
This has potentially important consequences for quantum control and quantum 
computation, in that it can be regarded as a strategy for performing {\sl 
fault-tolerant control}. Even though the effects of the environment make it 
impossible to retain control over arbitrary unitary evolutions of a 
quantum system, an effective dynamics can be still reliably constructed over a 
restricted set of transformations. 
 
The starting point of the method consists in recognizing that no relaxation
process can take place instantaneously. Accordingly, one should be able to 
interfere with the associated dynamics by inducing motions
into the systems, which are faster than the shortest time scale 
accessible to the reservoir degrees of freedom. The usage of tailored 
time-dependent perturbations as a tool to improve system performances has
a long history within high-resolution Nuclear Magnetic Resonance spectroscopy,
where versatile decoupling techniques are available to 
manipulate the overall spin Hamiltonian \cite{ernst}.
Despite this enlighting similarity, the construction of 
analogous procedures for open quantum systems faces an important conceptual 
difference, for we assume that any decoupling action can be exerted 
only on the system variables, the environment being contributed by a huge 
number of uncontrollable quantum degrees of freedom.

We consider a quantum system $S$ coupled to an arbitrary bath $B$, which 
together form a closed system defined on the Hilbert space 
${\cal H}={\cal H}_S \otimes {\cal H}_B$, ${\cal H}_{S}$ and
${\cal H}_{B}$ denoting $S$ and $B$ Hilbert spaces respectively. 
The overall Hamiltonian can be written in a concise form as
\begin{equation}
H_0=H_S \otimes \openone_B + \openone_S \otimes H_B + H_{SB}=
\sum_{\alpha} {\cal S}_\alpha \otimes {\cal B}_\alpha \;, 
\label{hamiltonian}
\end{equation}
where $\openone$ is the identity operator and the bath operators 
${\cal B}_\alpha$ are supposed to be linearly independent. 
Being $H_0$ Hermitian, the linear space spanned by the system operators $
{\cal S}_\alpha$ is a self-adjoint subspace in the vector space $B({\cal H}_S)$
of bounded operators acting on ${\cal H}_S$. We shall assume that the unwanted
noise-inducing Hamiltonian $H_{SB}$ is only contributed by a finite subset of 
open-system couplings. We call {\sl interaction space} ${\cal I}_S
\subseteq B({\cal H}_S)$ the corresponding finite dimensional subspace.
The second ingredient we introduce is the {\sl control algebra}, ${\cal C}_S$, 
which is generated by the repertoire of Hamiltonians we can turn on for $S$ 
to implement decoupling. 
We allow for possibilities where ${\cal I}_S \not= {\cal C}_S$. 

If $\rho_{tot}(0)=\rho_S(0) \otimes \rho_B(0)$ is the initial 
state over ${\cal H}$, the open-system evolution of $S$ is the
coarse-grained dynamics $\rho_S(0) \mapsto \rho_S(t)$ = $\mbox{Tr}_B\{ 
\rho_{tot}(t) \}$, $\mbox{Tr}_B$ denoting partial trace over 
${\cal H}_B$ \cite{gardiner}. The relaxation dynamics for $\rho_S(t)$, which 
involves a combination of quantum decoherence and dissipation mechanisms 
depending on the nature of the coupling operators, may display a complicate 
time dependence. In the simplest case, the off-diagonal matrix elements of
$\rho_S(t)$ behave like $\exp(-t/\tau_{rel})$, $\tau_{rel}$ indicating the 
time scale for significant departure from unitarity and irreversible loss
of quantum coherence.

In order to protect the evolution of $S$ against the effect of the
interaction $H_{SB}$, we start by seeking a perturbation $H_1(t) \in
{\cal C}_S$ to be added to $H_0$ as a suitable decoupling interaction, 
$H(t)= H_0 + H_1(t) \otimes \openone_{B}$.
We restrict here to situations where the
control field is {\sl cyclic}, {\it i.e.}, associated to a decoupling
operator $U_1(t)$ that is periodic over some cycle time $T_c >0$:
\begin{equation}
U_1(t)\equiv T\exp\bigg\{ \hspace{-0.5mm} -i \int_0^t\, du \,H_1(u)
\hspace{-0.5mm} \bigg\} = U_1(t+T_c) \;. 
\label{cyclicity}
\end{equation}
In the interaction representation associated with $H_1(t)$, defined by 
$\rho_{tot}(t)=U_1(t) \tilde{\rho}_{tot}(t) U_1^\dagger (t)$ on ${\cal H}$, 
time evolution is ruled by a transformed time-varying Hamiltonian,
\begin{equation}  
\tilde{H}(t) = U_1^\dagger (t) H_0 U_1(t) =
\sum_\alpha \Big[ U_1^\dagger (t) {\cal S}_\alpha U_1(t) \Big] \otimes 
{\cal B}_\alpha \;.    
\label{htilde}
\end{equation}
Since $U_1(T_c)= \openone_S$, the evolution in the original Schr\"odinger 
representation can be easily derived. One can prove that after $N$ cycles, 
$T_N=N T_c$, 
\begin{equation} 
U_{tot}(T_N) = \mbox{e}^{-i \overline{H} T_N} 
\;, \label{propagator}
\end{equation}
where the motion of the system under the time-dependent field $H_1(t)$
has been replaced by a {\sl stroboscopic} development under an {\sl effective}
so-called average Hamiltonian $\overline{H}$ \cite{ernst}. The calculation of 
$\overline{H}$ is performed on the basis of a standard Magnus expansion of the
time-ordered exponential defining the cycle propagator $U_{tot}(T_c) 
=\exp{ ( -i \overline{H} T_c) }$, 
\begin{equation}
T\exp \bigg\{ \hspace{-0.5mm} -i \int_0^{T_c} du 
\, \tilde{H}(u)  \hspace{-0.5mm} \bigg\} =
\mbox{e}^{-i[\, \overline{H}^{(0)} + \overline{H}^{(1)} + \ldots\,]\, T_c} \;, 
\label{magnus}
\end{equation}
where the various contributions collect terms of equal order in the 
transformed Hamiltonian. In particular, 
\begin{equation} 
\overline{H}^{(0)}  =  {1 \over T_c} \int_0^{T_c} du \, \tilde{H}(u)
\;. \label{zeroth} 
\end{equation}

We shall say that $k$th-order decoupling is achieved if the control field 
$H_1(t)$ can be devised so that contributions mixing $S$ and $B$ degrees of
freedom are no longer present in $\overline{H}^{(0)}$ and the first
nonvanishing correction arises from $\overline{H}^{(k)}$, $k \geq 1$. 
Owing to the fact that the cycle time $T_c$ enters the Magnus series as a
controllable expansion parameter, we examine the limit of {\sl fast} control,
where the lowest-order terms are expected to provide an accurate description 
({\sl first-order decoupling}).
Formally, for a finite evolution time $T$, this requires considering
$T_c=T/N$ in the limit as $N\rightarrow\infty$. The Magnus series 
defining evolution over a single cycle converges for sufficiently
large $N$ whenever $\overline{H}^{(r)}= O(T_c^r) = O(1/N^2)$ for $r\geq 2$.
As a result, in the limit of arbitrarily fast control, contributions higher 
than zeroth-order are negligible in (\ref{magnus}) and we can focus on the 
problem of designing the effective Hamiltonian $\overline{H}^{(0)}$. 

We now show that the time average defining $\overline{H}^{(0)}$ can
be made identical to a group-theoretical averaging procedure\cite{groups}. 
Since we can apply any Hamiltonian in the control algebra,
full control of the decoupling propagator $U_1(t)$ is available over 
the associated set of unitary transformations. 
Let ${\cal G}$ be any finite group of unitary operators that generates 
${\cal C}_S$, ${\cal G}\equiv \{ g_j \}$, $j=0, \ldots, |{\cal G}|-1$, 
$|{\cal G}|\equiv$ ord$({\cal G})$. Then the map
\begin{equation}
{\cal S} \mapsto \overline{\cal S} \equiv \Pi_{\cal C}({\cal S}) =
 {1 \over |{\cal G}| } 
\sum_{g_j \in {\cal G}} \, g_j^\dagger\, {\cal S} \, g_j \;, \hspace{5mm}
{\cal S} \in B({\cal H}_S)\;, 
\label{map}
\end{equation}
is the unique operation projecting on the so-called {\sl centralizer} of 
${\cal G}$. Equivalently, since averaged operators $\overline{\cal S}$ 
commute with every $g_j$, they belong to the so-called {\sl commutant} 
${\cal C}$ of the control algebra. Notice that ${\cal C}$ is closed under 
commutation. The map (\ref{map}) 
is implemented through a simple piecewise constant decoupling operator:
\begin{equation}
U_1(t) \equiv g_j\;, \hspace{5mm} j \,\Delta t \leq t < (j+1) \,\Delta t \;, 
\label{decoupling}
\end{equation}
corresponding to a partition of the cycle time $T_c$ into $|{\cal G}|$ 
intervals of equal length $\Delta t \equiv T_c /|{\cal G}|$. Then, by 
(\ref{htilde}), 
\begin{equation}
\overline{H}^{(0)} = \Pi_{\cal C} (H_0) = \sum_\alpha \, 
\overline{\cal S}_\alpha \otimes {\cal B}_\alpha \;, 
\label{proofa}
\end{equation}
which, by virtue of the quantum operation (\ref{map}), displays 
well-defined symmetry properties.

The decoupling prescription (\ref{decoupling}) requires the 
capability of instantaneously changing the evolution operator from $g_j$ 
to $g_{j+1}$ over successive subintervals, implying arbitrarily strong 
``kicks'' of control field. 
Such impulsive full-power control configurations correspond to 
so-called {\sl quantum bang-bang controls} as introduced in \cite{viola}. In 
fact, this method can be seen to provide an explicit control implementation 
of an abstract unitary symmetrization procedure recently proposed by Zanardi 
\cite{zanardi}.

Two different situations arise depending on the knowledge available
on the system-bath interaction $H_{SB}$. Let us first suppose that no 
knowledge is assumed, in which case the environmental coupling is 
completely arbitrary and ${\cal I}_S = B({\cal H}_S)$. We 
can prove the \vspace{1.5mm} following:
\par\noindent
{\it Theorem.} Let $S$ be a finite dimensional system and let the interaction 
with the environment be arbitrary, 
${\cal I}_S =B({\cal H}_S)$. Then, in the limit of arbitrarily fast control 
rate, the evolution of observables in the control algebra can be suppressed 
arbitrarily well:
\begin{equation}
\lim_{N \rightarrow \infty} \mbox{Tr}_S \{ A\,\rho_S(T=NT_c) \}  
 =  \mbox{Tr}_S\{ A\,\rho_S(0)\} \;. 
\label{thmi}  
\end{equation}
\par\noindent
If, in addition, the control algebra is maximal, ${\cal C}_S = B({\cal H}_S)$,
then complete first-order decoupling is achievable through system 
manipulations alone:
\begin{equation}
\lim_{N \rightarrow \infty} \rho_S(T=NT_c) = \rho_S(0) \;. 
\label{thmii}  
\end{equation}
\par\noindent
{\it Proof.} Let $A\in {\cal C}_S$. The first statement follows by 
Eq. (\ref{propagator}) with Hamiltonian (\ref{proofa}) and the fact that   
$[A,\overline{\cal S}_\alpha] =0$ for every $\alpha$. If ${\cal C}_S$ 
consist of all operators, then the commutant ${\cal C}$ only contains
$c$-numbers, $\overline{\cal S}_\alpha = \lambda_\alpha
\,\openone_S$ in (\ref{proofa}) and the result follows.
\hfill\vspace{1.5mm}$\Box$

The group able to average system operators into the commutant
of ${\cal C}_S$ can be chosen as a set of linearly independent unitary 
operators realizing a so-called {\sl unitary error basis} on ${\cal H}_S$ 
\cite{knill}. Such subgroups always exist for finite dimension. 
Within the class of piecewise constant decoupling sequences as considered 
above, it can then be shown that at least $\mbox{dim}({\cal C}_S)=
|{\cal G}|$ steps are needed in a cycle to attain decoupling. Since the 
effective system evolution is completely quenched by the decoupling 
procedure, we call this configuration {\sl maximal averaging}.

When some knowledge is available on the coupling $H_{SB}$, this information
can be exploited to engineer shorter decoupling sequences 
fulfilling specific symmetry constraints. For a given (known) interaction
space ${\cal I}_S$, 
the goal is to devise a control algebra able to selectively averaging out
${\cal I}_S$, $ \Pi_{\cal C}({\cal I}_S)=0$, while leaving 
invariant the sector in the operator space of the system containing some
useful dynamics. This requires that the error-inducing and the desired system 
operators in (\ref{hamiltonian}) transform according to different irreducible 
representations of ${\cal G}$. By using (\ref{proofa}), 
the effective open-system evolution over time $T$ is then governed by 
\begin{equation}
\lim_{N \rightarrow \infty} \rho_S(T=NT_c) = e^{-i \overline{H}_S T }
\,\rho_S(0)\, e^{+i \overline{H}_S T } \;, 
\label{select}  
\end{equation} 
the projected Hamiltonian $\overline{H}_S=\Pi_{\cal C}(H_S)$ only 
consisting of operators 
belonging to ${\cal C}$. Equivalently, {\sl the set of operators in the 
commutant
of ${\cal C}_S$ determines the reliable manipulations left 
available for effective system control.} For a given ${\cal I}_S$, the 
identification of a minimal group ${\cal G}$ able to produce decoupling is
nontrivial. We provide an illustrative example.

Let us discuss a $K$-qubits dissipative quantum register \cite{qcomp}. 
The maximum possible complexity of error 
generation arises in the presence of so-called {\sl total decoherence}, 
whereby combined errors can occur to any number of qubits. 
In this case, ${\cal I}_S = B({\cal H}_S)\simeq ({\sf C}^2)^{\otimes K}$ 
and maximal
averaging is demanded to decouple the register from quantum noise. 
Since an error basis on ${\sf C}^{2 \, \otimes K}$ is 
generated via the tensor product
of the standard Pauli bit/sign-flip error basis \cite{knill}, 
a possible choice is
${\cal G} = \{ \openone_S, \sigma_\alpha^{(i)}\}^{\otimes K}$, $\alpha=
x,y,z$, $i=1,\ldots,K$. Thus, a number of $|{\cal G}|=4^K$ steps is 
required for minimal length decoupling sequences. A more efficient 
averaging is possible if the relevant register-bath coupling is known to 
be {\sl linear} in single-qubit operators, $H_{SB}=
\sum_{i,\alpha} \, \sigma^{(i)}_\alpha \otimes {\cal B}^{(i)}_\alpha$. 
Under that condition, which is met for both independent decoherence 
(dim(${\cal I}_S)=3K$) and collective decoherence (dim(${\cal I}_S)=3$), 
selective averaging suffices to decouple from errors. One can show that the
tensor power of the Pauli group,
${\cal G} = \{ \openone_S, \otimes_{i=1}^K \sigma_\alpha^{(i)} \}$, 
$\alpha = x,y,z$, $|{\cal G}|=4$, represents a minimal choice. In terms of the 
control field $H_1(t)$, decoupling is then enacted by cycling the qubits 
in the register 
through sequences of {\sl collective} $\pi$-pulses along two axes,  
{\it e.g.}, $\Delta t - \pi_x - \Delta t - \pi_{-z} - \Delta t-
\pi_{-x} - \Delta t- \pi_{-z}$, $\Delta t= T_c/4$. In the special case 
of a purely decohering coupling, $H_{SB}=\sum_i \, \sigma_z^{(i)} 
\otimes {\cal B}_z^{(i)}$, the scheme can be further simplified by taking 
${\cal G}=\{ \openone_S, \otimes_{i=1}^K \sigma_x^{(i)} \}$, $|{\cal G}|=2$, 
which reproduces the situation analyzed in \cite{viola} for $K=1$. In the 
decoupling regime, the register effectively behaves as a noiseless {\sl 
quantum memory}, which is an essential 
ingredient for various quantum cryptographic and communication schemes 
\cite{qcomp,vaidman}. In addition, one can now perform in a fault-tolerant way 
any logical operation 
belonging to the commutant of the tensor power Pauli group, which is generated
through commutation by matrices of the form $\sigma^{(i)}_\alpha 
\sigma^{(j)}_\alpha $, $i,j \in \{1,\ldots,K\}$, $\alpha=x,y,z$.

The results discussed so far show that decoherence and decay can be 
completely suppressed in the limit of infinitesimally short control time 
scale $\Delta t \rightarrow 0$.
In order to convert this mathematical limit into a physically meaningful 
condition, we recall that complete information about the 
fluctuation-dissipation properties of a macroscopic bath is 
encapsulated in the spectral density function $J(\omega)$, 
measuring the density of modes at frequency 
$\omega$ multiplied by the square of the system-mode 
coupling strength. Quite generally, relaxation rates $\gamma =
\tau_{rel}^{-1}$ arise from an integration over the reservoir modes of the
effects due to thermal and 
vacuum fluctuations, weighted with $J(\omega)$. This integration does not 
extend to arbitrarily large frequencies. For every physical spectral 
density 
function, a finite ultraviolet cut-off frequency $\omega_c$ always exists, 
such that $J(\omega) \rightarrow 0$ for $\omega > \omega_c$. 
The characteristic {\sl memory time} $\tau_c \sim \omega_c^{-1}$, which is
often set to be zero as a part of Markov approximation \cite{gardiner}, 
determines the fastest 
time scale accessible to reservoir correlations. The condition 
$\Delta t \ll \tau_c$ replaces, in a realistic scenario, the 
ideal limit $\Delta t \rightarrow 0$. Thus, the physical requirement 
for decoupling is $\Delta t \lesssim \tau_c$ or 
$ \omega_c \, \Delta t \lesssim 1$, {\it i.e.}, {\sl the motion 
induced by the decoupling field needs to be faster than the fastest time 
scale characterizing the unwanted interactions}. 

In the presence of a small but finite decoupling time scale $\Delta t$,
the representation of the average Hamiltonian $\overline{H}$ in terms of
the lowest-order contribution $\overline{H}^{(0)}$ is approximate 
due to the higher-order terms. Consequently, decoupling is 
imperfect and relaxation dynamics still occurs even in the presence of
control. For first-order decoupling, the leading correction is 
due to $\overline{H}^{(1)}$ and the key
observation to estimate the decoupling accuracy as a 
function of $\Delta t$ is that the overall coupling
strength to the bath has been renormalized by a factor of the 
order $\omega_c \Delta t$. Since the spectral density 
depends quadratically upon the interaction strength, {\sl all the relevant 
relaxation effects tend to be suppressed by a factor of the order 
$(\omega_c \Delta t)^2$}. For a generic $k$th-order decoupling scheme, the 
controlled relaxation rate is able to be reduced as
\begin{equation}
{\gamma^C \over \gamma } \approx 
( \omega_c \, \Delta t )^{2k} =
\bigg( {\Delta t \over \tau_c} \bigg)^{2k} \;, \hspace{5mm} k \geq 1 
\;.  \label{accuracy}
\end{equation}
Second-order decoupling can be realized via so-called 
symmetric cycles, whereby 
$U_1(T_c-t)=U_1(t)$. Being $\overline{H}^{(r)}=0$ for $r$ odd, the 
error is improved to $O(\Delta t/ \tau_c)^4$. Iterative pulse sequences 
for $k$th-order decoupling can be designed for specific systems like 
quantum registers.

In practice, the feasibility of the approach depends 
on both the relevant correlation times and the 
sophistication of the technology available to manipulate the specific 
physical system. In pulsed-NMR experiments \cite{ernst}, where 
relaxation mechanisms due to ``slow'' nuclear motions may involve correlation
times longer than $10^{-8}$s, the main current limitation is represented by 
the pulse duration, $\tau_P \approx 1\mu$s. In atomic physics,
decoupling methods could prove to be viable for damped harmonic oscillators
schematizing the vibrational motion of trapped ions, since relevant cut-off 
frequencies may be estimated around 100 MHz and a variety of experimental 
techniques exists for coherent optical manipulation \cite{eng}. 
As another potential area of applications, we mention semiconductor-based 
structures.  Here, correlation times around $\omega_{Debye}^{-1} \approx 
10^{-13}$ s are comparable to the sub-ps time scale where control 
operations have been demonstrated \cite{heberle} 
and longer than the femtosecond scale of modern ultrafast laser 
technology. Rapid advancements in the capabilities of coherent control gives 
hope that, if not within the reach of present technology, 
implementations of decoupling schemes can be envisaged in a close future. 
In particular, since quantum computing resources are still 
a stringent practical requirement, decoupling
techniques could be valuable compared to conventional 
error-correction networks in the field of NMR, ion-trap or solid state quantum 
computation.           
    
In summary, we showed how to manipulate the irreversible 
component of open-system evolutions through the application of
external controllable interactions. Maximal and selective 
decoupling were introduced within a common group-theoretical framework and 
their relevance to the issue of designing controlled effective open-system
evolutions elucidated.  In the spirit of 
weakening the decoupling requirements as much as possible, the main 
question raised by the present analysis concerns the characterization
of {\sl fault-tolerant} decoupling schemes or, equivalently,  
the degree of decoupling attainable in the presence of power-limited and 
imperfect control operations. Work in ongoing along these directions.

This work was supported by ONR, AFOSR, and DARPA/ARO under the 
Quantum Information and Computation initiative and the NMR 
Quantum Computing initiative. 
E. K. received support from the NSA.

\end{multicols}

\end{document}